\documentclass[12pt]{article}
\usepackage{epsfig}
\usepackage{float}
\usepackage{perpage}
\MakeSorted{figure}
\MakeSorted{table}
\usepackage[section]{placeins}
\usepackage{lipsum}
\usepackage{url}
\usepackage{titling}
\usepackage[noadjust]{cite}
\usepackage{filecontents}

\def\beq{\begin{equation}}
\def\eeq#1{\label{#1}\end{equation}}
\def\eeqn{\end{equation}}

\def\beqa{\begin{eqnarray}}
\def\eeqa#1{\label{#1}\end{eqnarray}}
\def\eeqan{\end{eqnarray}}

\let\bar=\overbar

\def\Dslash{\not{\hbox{\kern-4pt $D$}}}
\def\dslash{\not{\hbox{\kern-2pt $\del$}}}

\def\msb{{\bar{\ssstyle M \kern -1pt S}}}

\date{}
\begin{document}

\title{Novel Application of Density Estimation Techniques in Muon Ionization Cooling Experiment~\thanks{Talk presented at the APS Division of Particles 
and Fields Meeting (DPF 2017), July 31-August 4, 2017, Fermilab. C170731}
}

\bigskip\bigskip

\author{Tanaz Angelina Mohayai, Pavel Snopok,\\
Illinois Institute of Technology, Chicago, IL, USA, \\
David Neuffer, \\
Fermilab, Batavia, IL, USA, \\
Chris Rogers, \\ 
STFC Rutherford Appleton Laboratory, Didcot, Oxfordshire, UK, \\
for the MICE Collaboration}
\maketitle

\abstractname{}

The international Muon Ionization Cooling Experiment (MICE) aims to demonstrate muon beam ionization cooling for the first time and constitutes a key part of the R\&D towards a future neutrino factory or muon collider. Beam cooling reduces the size of the phase space volume occupied by the beam. Non-parametric density estimation techniques allow very precise calculation of the muon beam phase-space density and its increase as a result of cooling. These density estimation techniques are investigated in this paper and applied in order to estimate the reduction in muon beam size in MICE under various conditions.
		

\section{Muon Ionization Cooling Experiment}
The Muon Ionization Cooling Experiment (MICE) is an accelerator-physics experiment located at the Rutherford Appleton Laboratory (RAL) in Oxfordshire, United Kingdom. Its aim is to demonstrate muon beam size reduction (``cooling'') by passing a beam of muons through absorbing material under suitable beam optics conditions. When muons are produced from pion decay, they occupy a large phase-space volume. To fit them into cost-effective apertures of a future neutrino factory~\cite{ref1} or muon collider~\cite{ref2}, the beam size needs to be reduced. Ionization cooling is the only technique capable of cooling a beam of the required intensity within the short muon lifetime. A cooling factor of almost one million has been achieved in simulation~\cite{ref3}.

\begin{figure}[htb]
\begin{center}
\advance\leftskip-3cm
\advance\rightskip-3cm
\includegraphics[keepaspectratio=true,scale=0.5]{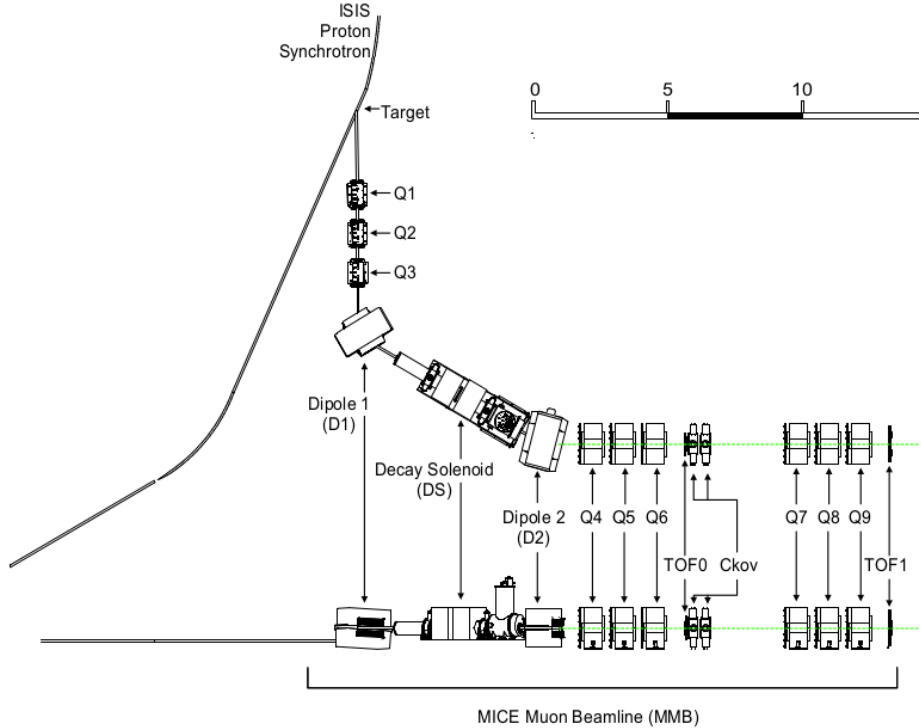}
\caption{Schematic diagrams (plan and elevation) of the MICE muon beamline~\cite{ref5}.}
\label{fig:beamline}
\end{center}
\end{figure}

In the MICE beamline (Fig.~\ref{fig:beamline}), pions are produced from the interaction of a titanium target with the ISIS proton beam~\cite{ref4}, focused by a triplet of quadrupole magnets (Q1--Q3) and momentum selected by a dipole magnet (D1). The pion-decay muons are also momentum selected by a dipole magnet (D2) and focused by a pair of triplet quadrupole magnets (Q4--Q9)~\cite{ref5}. One of the parameters that can be varied in MICE is the input muon momentum. The recently taken MICE data has input muon momentum of $140$ MeV/$c$ (ISIS Cycle $2016$/$04$, Run Setting $1.2$ in~\cite{ref6}).

\begin{figure}[htb]
\begin{center}
\advance\leftskip-3cm
\advance\rightskip-3cm
\includegraphics[keepaspectratio=true,scale=0.5]{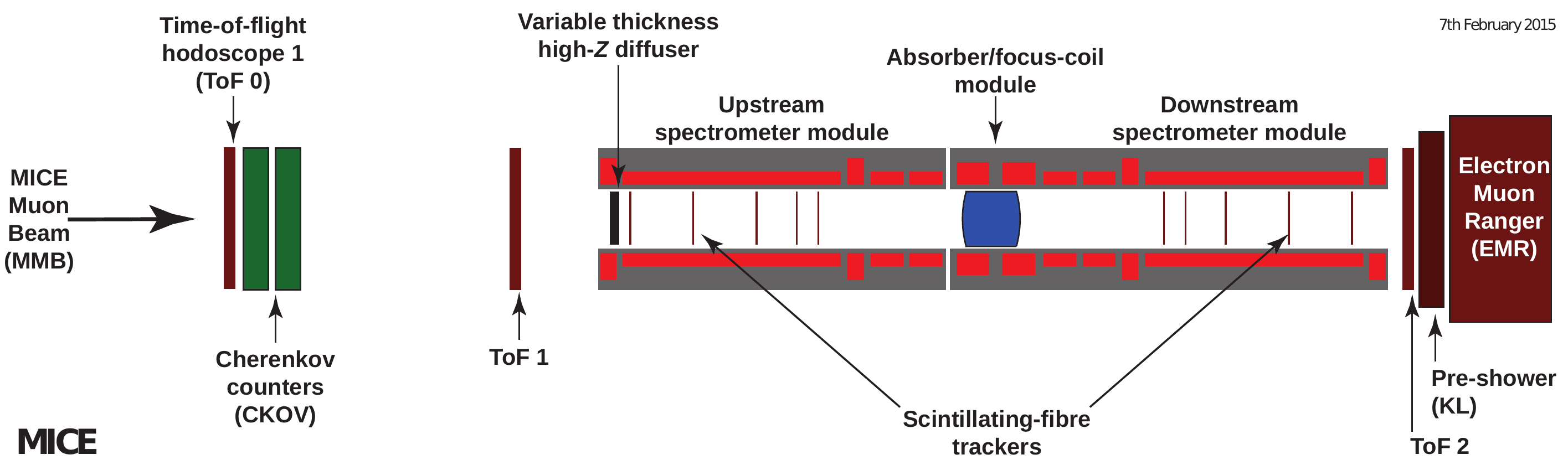}
\caption{Schematic diagram of the Muon Ionization Cooling Experiment (MICE) in its current Step IV configuration, with upstream and downstream spectrometers surrounding a cooling cell, and particle identification detectors (ToF, Cherenkov, KL, and EMR)~\cite{ref5}.}
\label{fig:mice}
\end{center}
\end{figure}

The MICE muon beam at the entrance to the experiment has a slight pion and electron contamination. To ensure muon beam purity, MICE uses a series of particle identification (PID) detectors as shown in Fig.~\ref{fig:mice}. These PID detectors efficiently tag contaminating particles, in particular, pions, allowing them to be rejected from the muon cooling measurements; the residual pion contamination is less than $\sim$1\%~\cite{ref7}. Pion rejection at the entrance of the experiment is particularly important for phase-space density studies, as pions left in the beam would decay in the MICE cooling channel, affecting the increase in the muon beam density (increase in phase-space density signifies beam cooling).

Ionization cooling in MICE occurs when muons lose momentum as a result of interactions with an absorbing material. This process is measured via a pair of tracking detectors~\cite{ref8} upstream and downstream of the MICE absorber. Each tracker is immersed in the solenoidal fields of the Spectrometer Solenoid (SS) magnets, and the absorber is placed within a separate focusing solenoidal magnet (Absorber Focus Coil, or AFC module). For measurement of beam cooling in MICE, the phase-space density and volume of the input and output beam distributions at the tracker reference planes (tracker planes closest to the absorber) are compared. An increase in phase-space density and decrease in phase-space volume for a surface enclosing a fixed percentage of the beam particles signifies beam cooling. In the current configuration of the experiment, the aim is to demonstrate muon beam cooling using lithium hydride (LiH) and liquid H$_{2}$ (LH$_{2}$) absorbers. 

To study the muon ionization cooling process in detail, a good understanding of the ionization energy loss~\cite{ref9} and the multiple Coulomb scattering of muons in material~\cite{ref10} are required. One way to describe the beam cooling process is to use the rate of change of the normalized transverse RMS emittance~\cite{ref11}: 
\begin{equation}
\frac{d\varepsilon_{\perp}}{ds} \approx -\frac{\varepsilon_{\perp} }{\beta ^{2}E_{\mu }}\left \langle \frac{dE}{ds} \right \rangle +\frac{\beta _{\perp} (13.6\textnormal{MeV}/c)^{2}} {2\beta ^{3}E_{\mu }m_{\mu }X_{0}},
\label{eq:cooling}
\end{equation}
where $E_{\mu}$ is the muon energy, $\beta c$ the muon velocity, $dE/ds$ the magnitude of the ionization energy loss, $m_{\mu}$ the muon mass, $X_{0}$ the radiation length of the absorber material, and $\beta_{\perp}$ the transverse beta function at the absorber. 

The first term in Eq.~(\ref{eq:cooling}) represents beam cooling from ionization energy loss, and the second describes beam heating from multiple Coulomb scattering. The minimum achievable emittance, or equilibrium emittance, for a given material and focusing conditions, is obtained by setting the rate of change of the normalized transverse RMS emittance to zero:
\[
\varepsilon _{\perp } \cong \frac{\beta_{\perp }(13.6\textnormal{MeV})^{2} }{2X_{0}\beta m_{\mu }}\left \langle \frac{dE}{ds} \right \rangle^{-1}.
\]
A smaller equilibrium emittance is achieved when the beta function, $\beta_{\perp}$, is minimized, and $X_{0}$ maximized, for a given energy loss rate, $dE/ds$~\cite{ref11}. In MICE, a small beta function is achieved by focusing the beam tightly at the absorber using the AFC, and a large radiation length is achieved using low-$Z$ absorbing materials such as LiH and LH$_{2}$. In addition, the input emittance in MICE can be varied using a diffuser (a high-$Z$, tungsten or brass insert) located at the entrance of the upstream tracker (Fig.~\ref{fig:mice}). The larger the input emittance compared to the equilibrium emittance, the larger the beam cooling or increase in phase-space density. The simulations discussed in Sec.~\ref{sec:density_in_mice} were carried out for muon beams of $6\pi$~mm$\cdot$rad and $10\pi$~mm$\cdot$rad RMS input emittances with an equilibrium emittance of approximately $5\pi$~mm$\cdot$rad.

In general, the change in emittance is expected to be small for a short channel such as MICE. In addition, because of the nonlinear effects (for example, the particle distribution is not Gaussian, especially downstream of the absorber), the RMS emittance does not accurately describe the volume of phase space occupied by the beam. Therefore, a better estimate of the phase-space volume than RMS emittance is needed. Density estimation (DE) techniques provide such an estimate. 

\section{Non-Parametric~Density~Estimators}
As already mentioned, one way to measure the volume of phase space occupied by a beam is to compute its root-mean-square (RMS) emittance. This approach works well for Gaussian distributions. When the beam distribution is non-Gaussian (e.g., in the presence of nonlinear effects), the RMS emittance is not a precise measure of the volume occupied by the beam. In this case, the phase-space volume of the beam can be studied using density estimation (DE) techniques. 

There are two DE approaches, parametric and non-parametric. The parametric approach makes an assumption about the underlying phase-space density or probability density function (PDF) and estimates the distribution parameters. The RMS emittance measurement is an example of a parametric method. Non-parametric methods make no assumptions about the underlying particle distribution, and the individual particles within the beam are allowed to ``speak for themselves'' in determining the beam distribution. Non-parametric DE techniques typically rely on a smoothing technique to estimate the density (e.g., kernels, series, or splines), and the level of smoothing is generally tuned by a parameter.

Consider a one-dimensional sample distribution of size $N$ divided into bins of width $h$. The lowest bin edge is at $x_{0}$ (typically, the smallest value in the sample). Then with $n_m$ being the number of data points that fall within the $m^{th}$ bin, one of the most widely-used and oldest DE techniques, the histogram estimator, estimates the PDF as
\[
\hat{f}(x)=\frac{n_m}{Nh},
\]
where $\hat{f}$ is the estimated density and $[x_{0}+mh < x < x_{0}+(m+1)h]$. The histogram estimator has several challenges. In particular, the choice of the bin edge can affect the estimated density. This is a problem in multiple dimensions where the bin edge needs to be determined for each dimension separately~\cite{ref12}.

Another widely studied DE technique is kernel density estimation (KDE)~\cite{ref13}. The KDE technique uses kernel functions (smooth weight functions)~\cite{ref12}. To estimate the density everywhere in phase space, kernels are centered at each data point and are summed: 
\[
\hat{f}(\vec{r}) = \frac{1}{Nh^{q}\sqrt{(2\pi )^{q}}}\sum_{i=1}^{N}K\left(\frac{\vec{r}-\vec{R_{i}}}{h}\right),
\]  
where $\vec{r}$ is the reference point (e.g.,~data point or grid point) at which the density is being estimated, $K\left(\frac{\vec{r}-\vec{R_{i}}}{h}\right)$ is the kernel function centered at the $i^{th}$ data point (whose transverse coordinates are $\vec{R_{i}}$), $h$ is the parameter that tunes the level of smoothing of the density curve (known as the bandwidth parameter), $N$ is the total sample size, and $q$ is the dimensionality of the phase space (dimensionality in MICE is $4$). The kernel functions can be of Gaussian form,
\[
K\left(\frac{\vec{r}-\vec{R_{i}}}{h} \right) = \frac{1}{(2\pi)^{\frac{q}{2}}} \textnormal{exp}\left [- \frac{\left ( \vec{r}-\vec{R_{i}} \right)^{2}}{2h^{2}} \right].
\] 
Compared with the histogram approach, the density curve estimated using KDE is smoother, leading to a more straightforward data interpretation. In addition, unlike the histogram approach (where only the data points within the bins count), the kernels are centered at each data point, causing all data points to contribute to the density at the reference point $\vec{r}$. 

\begin{figure}[htb]
\begin{center}
\advance\leftskip-3cm
\advance\rightskip-3cm
\includegraphics[keepaspectratio=true,scale=0.5]{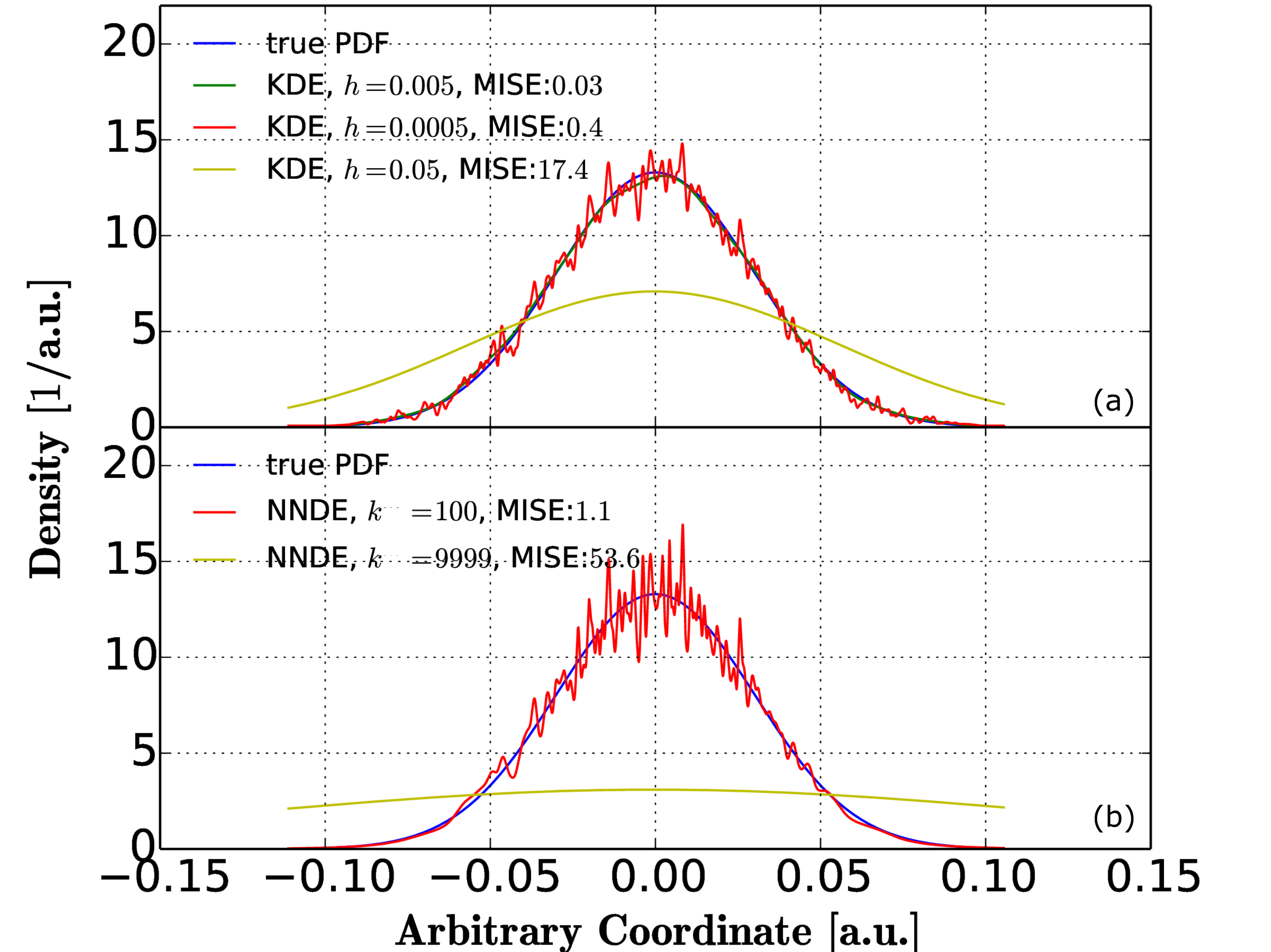}
\caption{Illustration of the bias$-$variance trade-off in a) KDE (upper plot) and b) NNDE (lower plot). The DE approximations are based on the sample size of $N=10,000$. The optimal bandwidth parameter, $h=0.005$ (green in the upper plot) reveals a density curve close to the true Gaussian PDF (blue in both plots) and if $h$ is decreased (red in the upper plot), the density curve gets noisier. The optimal $k$ parameter, $k=\sqrt{N}$ (red in the lower plot) reveals a curve that is noisy but still follows the true curve (blue) closely. Increasing these optimal values leads to an overly smooth curve (yellow in both plots).}
\label{fig:tradeoff}
\end{center}
\end{figure}

In KDE, the level of smoothing is generally tuned using the bandwidth parameter, $h$.~There are various ways of finding the optimal bandwidth parameter. One common choice is the bandwidth parameter that minimizes the discrepancy between the true (predicted) density and the estimated density. The commonly used measure of such discrepancy is the mean integrated square error (MISE)~\cite{ref12}, 
\[
\textnormal{MISE} = \left \langle \int \left \{ \hat{f}(\vec{r}) - f(\vec{r})\right \}^2   d\vec{r} \right \rangle, 
\]
where $\hat{f}(\vec{r})$ and $f(\vec{r})$ represent the estimated density and the true density. All DE techniques share an intrinsic property known as ``bias$-$variance trade-off'': if an attempt is made to reduce the bias or level of smoothing ($\propto h$) by selecting a small bandwidth parameter, $h$, the variance or level of noise ($\propto 1/h$) increases. This is illustrated in Fig.~\ref{fig:tradeoff}, where the true PDF is a Gaussian. The optimal bandwidth parameter ($h=0.005$ in Fig.~\ref{fig:tradeoff}a) leads to a density estimate that agrees well with the true density and has the smallest MISE.
The optimal bandwidth that minimizes the MISE for a true Gaussian or normal distribution is referred to as the normal reference bandwidth~\cite{ref12},
\[
h=\left(\frac{4}{q+2}\right)^{\frac{1}{q+4}}N^{-\frac{1}{q+4}}\Sigma,
\]
where $N$ is the sample size ($N=10,000$ in Fig.~\ref{fig:tradeoff}, $q$ is the phase-space dimensionality, and $\Sigma$ is the covariance matrix representing the correlations among the coordinates (the position and momentum phase-space coordinates of muons in MICE).

$k^{th}$ Nearest Neighbor Density Estimation (NNDE) is an alternative DE technique that uses the distance between a data point and its neighbor as the width of the kernel:
\[
\hat{f}(\vec{r}) = \frac{1}{Nd^{q}\sqrt{(2\pi )^{q}}}\sum_{i=1}^{N}K\left(\frac{\vec{r}-\vec{R_{i}}}{d}\right),
\]
where $d$ represents the distance between the $i^{th}$ data point with coordinate $\vec{r_{i}}$ and its $k^{th}$ nearest neighbor. This $k$ parameter (the ``order'' of the nearest neighbor) affects the level of smoothing in the estimated density curve. The optimal value is typically the square root of the sample size; as shown in Fig.~\ref{fig:tradeoff}b, this choice of $k$ ($k=100$) still results in an undesirably noisy curve. If the farthest neighbor is selected in an attempt to overcome the noise, the density curve becomes overly smooth. In addition, since in the tail of the distribution (beyond $1\sigma$), data points are generally more sparse than in the core, the NNDE technique could lead to a highly suppressed tail and a highly emphasized core. For these reasons, only the KDE technique has been applied to MICE simulation studies, with the possibility of extending the analysis to a technique which is a hybrid of NNDE and KDE. 

\begin{figure}[htb]
\begin{center}
\includegraphics[keepaspectratio=true,scale=0.6]{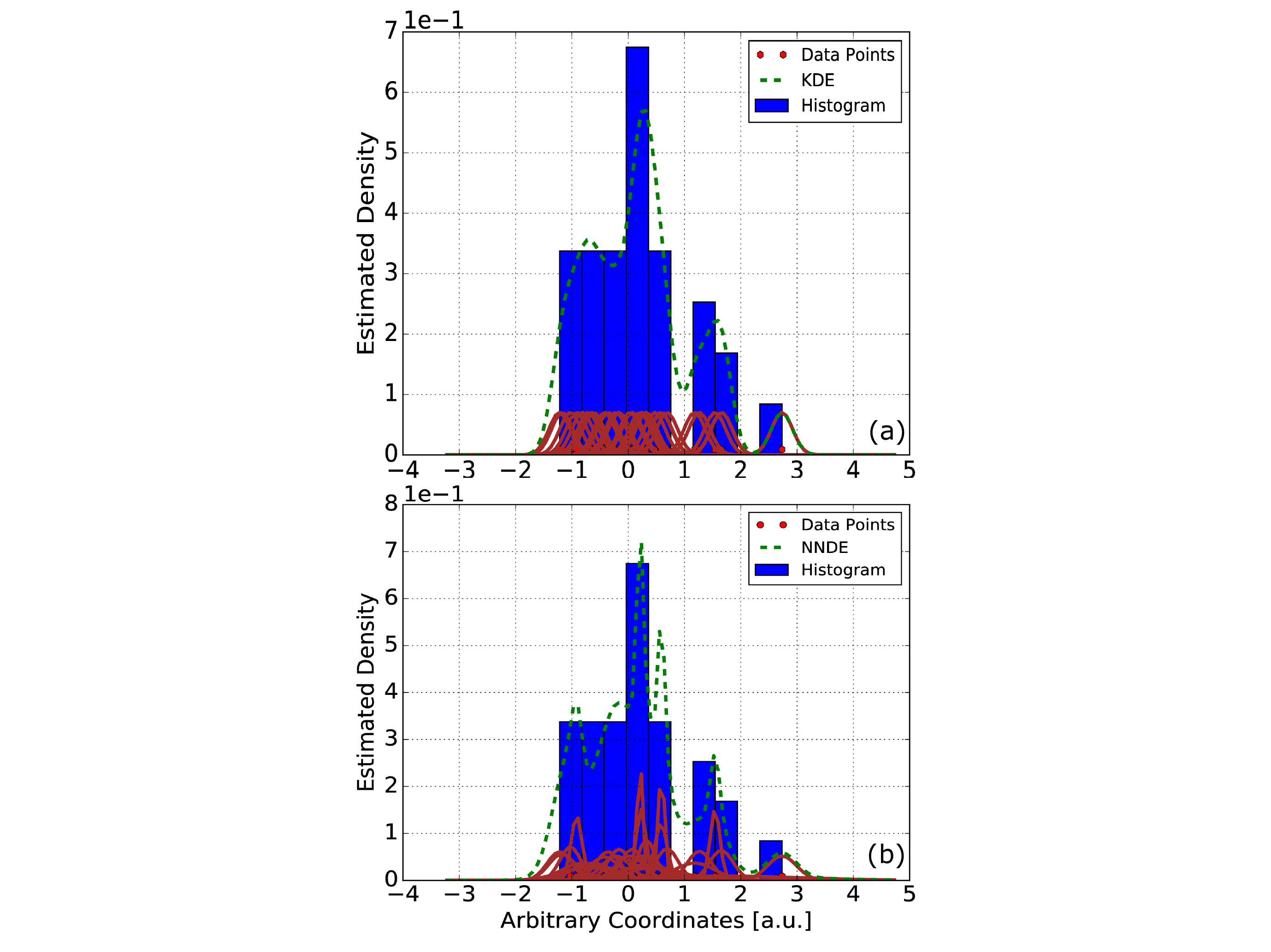}
\caption{Comparison of a) KDE (upper plot), b) NNDE (lower plot), and histogram DE (both). The individual kernel functions are shown in solid brown. The number of data points is $50$ and the true density is Gaussian.}
\label{fig:DE}
\end{center}
\end{figure}

Figure~\ref{fig:DE} compares the KDE (Fig.~\ref{fig:DE}a), NNDE (Fig.~\ref{fig:DE}b), and the histogram estimator (both). The KDE kernels in Fig.~\ref{fig:DE}a (solid brown) have fixed widths everywhere whereas the NNDE kernel widths in Fig.~\ref{fig:DE}b vary depending on the distances between data points. The KDE density curve (dashed green in Fig.~\ref{fig:DE}a) is smoother than the histogram and less noisy than the NNDE curve (dashed green in Fig.~\ref{fig:DE}b). In addition, compared with the KDE and NNDE curves, the histogram has a more step-wise nature. As a result, the KDE and NNDE curves better represent the underlying Gaussian distribution (the true density is Gaussian) than does the histogram estimator.

\section{Density Estimation in MICE}
\label{sec:density_in_mice}
The MICE simulation studies are done using the MICE Analysis User Software (MAUS v3.0.0) \cite{ref14}, XBOA~\cite{ref15}, and G4beamline v2.16~\cite{ref16}. MAUS routine generates the input muon distributions. XBOA converts the input beam to BLTrackfile format for tracking in G4beamline where MICE cooling channel geometry is defined.~The input beams of $10,000$ muons have $140$~MeV/$c$ reference momentum and transverse normalized input emittances of $6\pi$~mm$\cdot$rad (referred to as $6$--$140$) and $10\pi$~mm$\cdot$rad (referred to as $10$--$140$). They are selected to be Gaussian and are matched to the $3$~T field of the upstream spectrometer solenoid (SS). The magnet currents in the SS modules in G4beamline are the same as in the recent MICE runs~\cite{ref6}. Two of the five coils in the downstream SS (Match 1 and Match 2) are off. This leads to particle loss due to scraping and optical mismatch after passing through the LiH absorber. The transmission efficiency of the G4beamline simulated lattices is about $85$\% for the $6$--$140$ input beam and about $60$\% for the $10$--$140$ beam. No particle selection is applied downstream of the LiH absorber to discard scraped muons from the upstream sample. 

\begin{figure}[!htb]
\begin{center}
\advance\leftskip-3cm
\advance\rightskip-3cm
\includegraphics[keepaspectratio=true,scale=0.6]{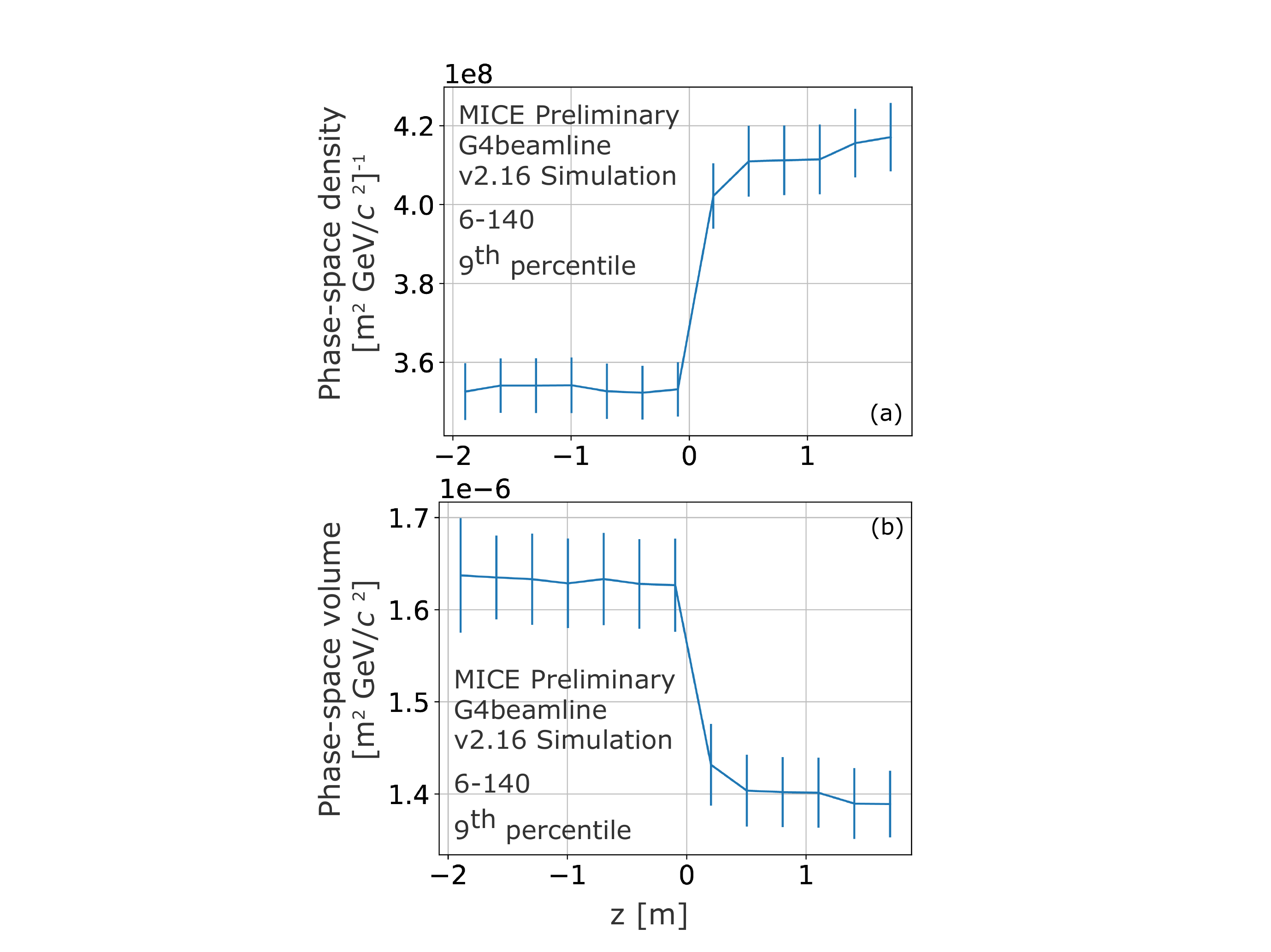}
\caption{a) Phase-space density (upper plot) and b) volume (lower plot) evolution for a 6--140 input beam. The LiH absorber is centered at $z=0$ m. The upstream and downstream tracker reference planes are located at $z=-1.9$ and $z=1.9$~m, respectively. A $20$\% increase in phase-space density and $15$\% reduction in phase-space volume is observed.}
\label{fig:6-140}
\end{center}
\end{figure}

\begin{figure}[!htb]
\begin{center}
\advance\leftskip-3cm
\advance\rightskip-3cm
\includegraphics[keepaspectratio=true,scale=0.6]{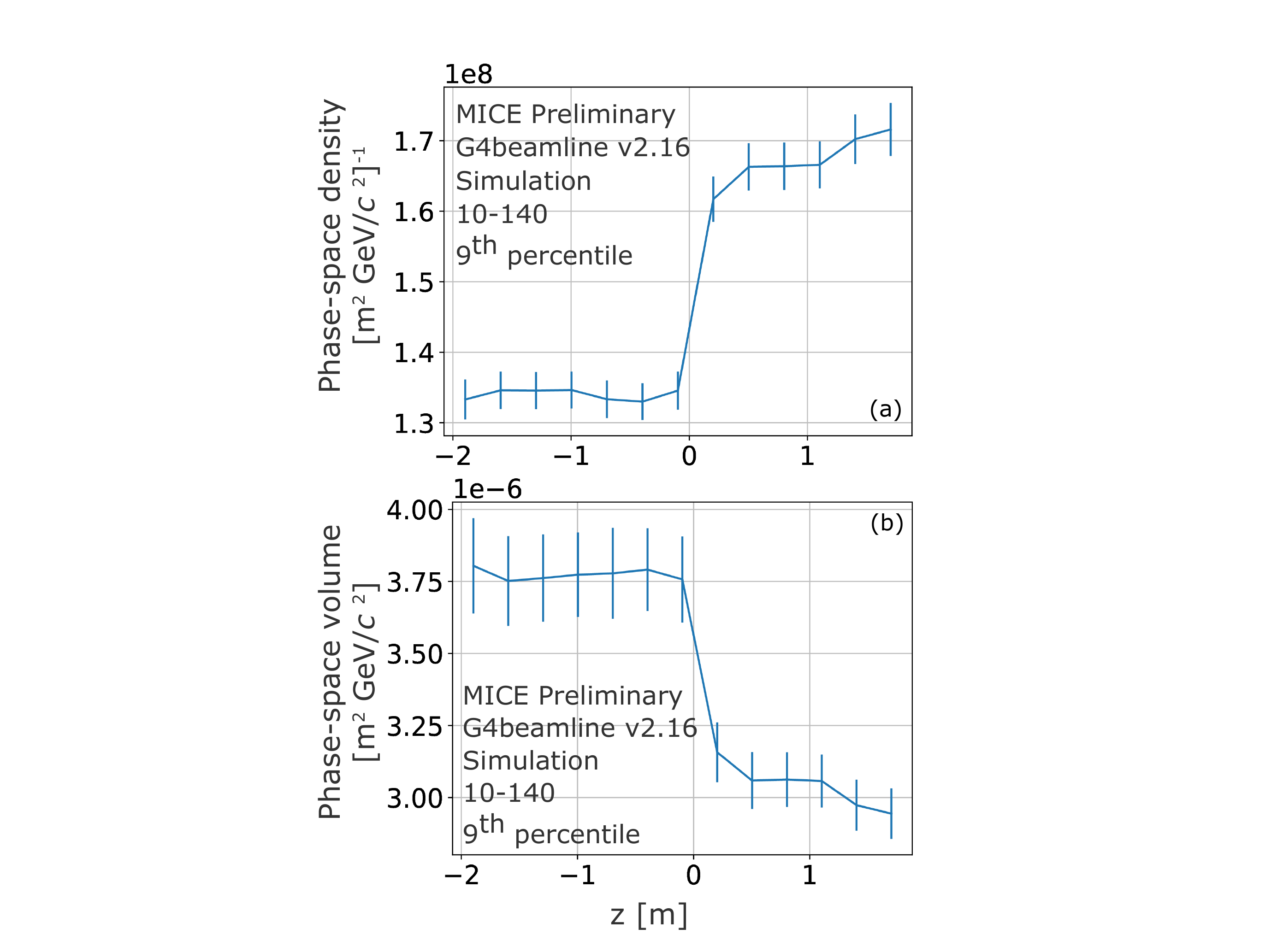}
\caption{a) Phase-space density (upper plot) and b) volume (lower plot) evolution for a 10--140 input beam. The LiH absorber is centered at $z=0$ m. The upstream and downstream tracker reference planes are located at $z=-1.9$ and $z=1.9$~m, respectively. A $28$\% increase in phase-space density and $22$\% reduction in phase-space volume is observed.}
\label{fig:10-140}
\end{center}
\end{figure}

Figures~\ref{fig:6-140} and~\ref{fig:10-140} show the evolution of the phase-space density and volume corresponding to the $9^{th}$-percentile contour or the beam core (the contour enclosing a volume containing $9$\% of the muon sample) along the MICE cooling channel with a 65-mm LiH absorber for $6$--$140$ and $10$--$140$ input beam settings. The KDE algorithm uses the transverse phase-space coordinates recorded by the G4beamline virtual detectors spaced uniformly along the $z$ axis from the upstream tracker reference plane at $z=-1.9$~m to the downstream tracker reference plane at $z=1.9$~m. The absorber is centered at $z=0$. An increase in phase-space density and reduction in phase-space volume in the beam core at the downstream tracker reference plane indicate cooling. The volume of the $9^{th}$-percentile contour is obtained using a Monte Carlo (MC) method, where MC points are generated randomly inside a bounding box for the contour. The volume enclosed by the contour is then determined as the volume of the box multiplied by the fraction of the MC points within the contour\cite{ref17,ref18,ref19,ref20}. The slight fluctuations in the volume curves are therefore statistical and expected. The error bars are the standard deviations of the repeated density and volume measurements (based on $10$ runs with $10,000$ muons each).  

\begin{figure}[!htb]
\begin{center}
\advance\leftskip-3cm
\advance\rightskip-3cm
\includegraphics[keepaspectratio=true,scale=0.6]{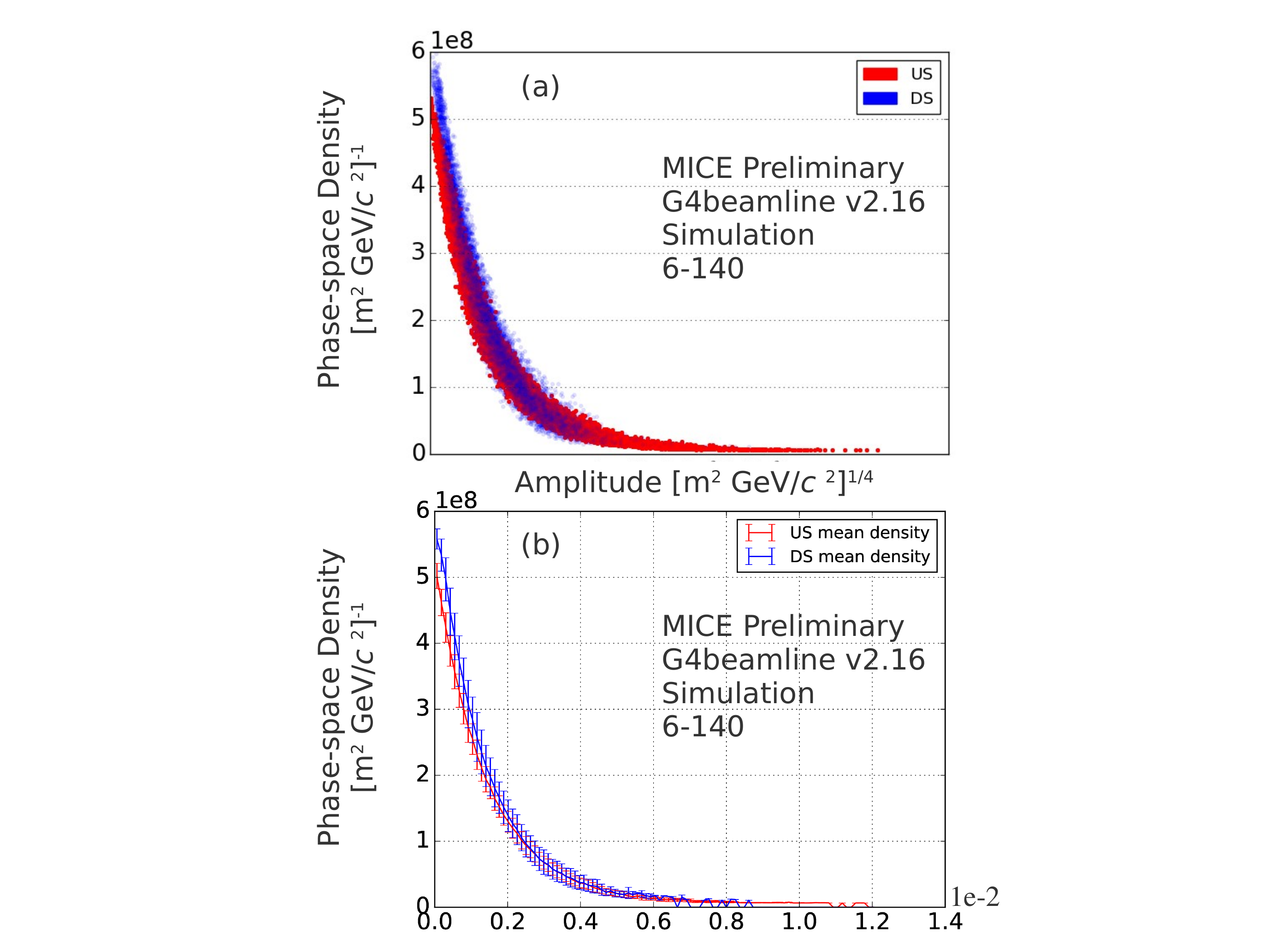}
\caption{a) Scatter plot of phase-space density versus beam amplitude for the $6$-$140$ input beam setting (upper plot) and b) average phase-space density versus beam amplitude (lower plot). Phase-space density increases closer to the core of the beam (smaller amplitude values) as a result of beam cooling, from upstream of the LiH absorber to downstream.}
\label{fig:amplitude-6-140}
\end{center}
\end{figure}

\begin{figure}[!htb]
\begin{center}
\advance\leftskip-3cm
\advance\rightskip-3cm
\includegraphics[keepaspectratio=true,scale=0.6]{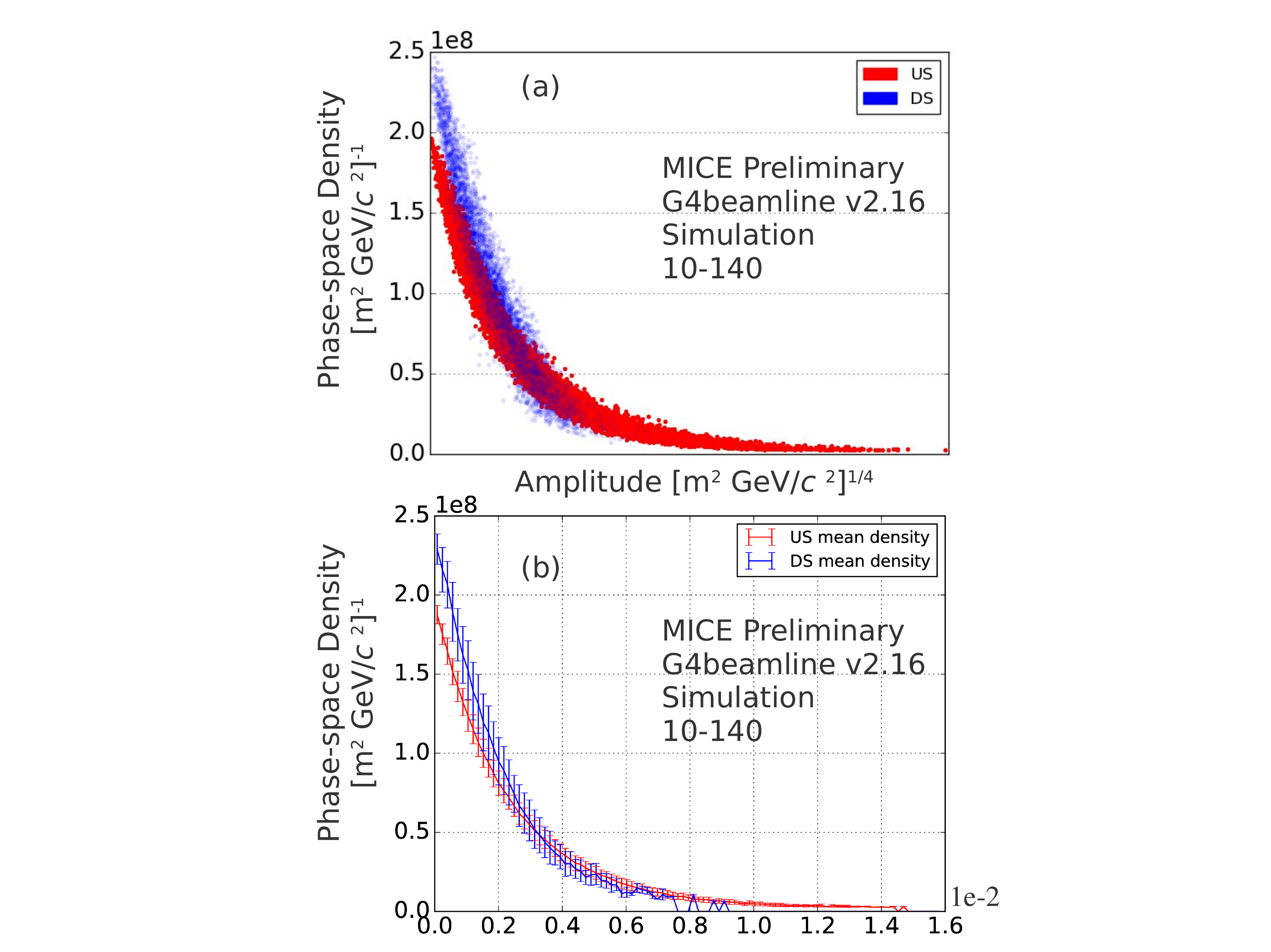}
\caption{a) Scatter plot of phase-space density versus beam amplitude for the $10$-$140$ input beam setting (upper plot) and b) average phase-space density versus beam amplitude (lower plot). Phase-space density increases closer to the core of the beam (smaller amplitude values) as a result of beam cooling, from upstream of the LiH absorber to downstream.~Average phase-space density versus beam amplitude is displayed in the lower plot.}
\label{fig:amplitude-10-140}
\end{center}
\end{figure}

Figures~\ref{fig:amplitude-6-140} and~\ref{fig:amplitude-10-140} display the increase in density at smaller beam amplitudes for $6$--$140$ and $10$--$140$ input beam settings. The transverse beam amplitude is obtained using the formula
\begin{equation}\label{eq:units}
A_{\perp}=\varepsilon_{\perp}\vec{r} ^{T} \Sigma\vec{r},
\end{equation}
where $\varepsilon_\perp$ is the normalized RMS emittance, $\Sigma$ is the covariance matrix, and $\vec{r}$ is the transverse phase space vector ($x$, $p_x$, $y$, $p_y$). Figures~\ref{fig:amplitude-6-140}a and~\ref{fig:amplitude-10-140}a are scatter plots of the KDE-based phase-space density estimate versus amplitude for the $6$--$140$ and the $10$--$140$ beam settings, respectively. The distribution is broad because the hyperellipsoid of constant amplitude is only an approximation to the surface of constant phase-space density. Figures~\ref{fig:amplitude-6-140}b and~\ref{fig:amplitude-10-140}b are obtained by binning the amplitudes in the scatter plots (Figures~\ref{fig:amplitude-6-140}a and~\ref{fig:amplitude-10-140}a) uniformly and averaging over the density values within each bin. The error bars are the standard deviations of the density values within each bin. As previously described, an increase in phase-space density indicates cooling, and we observe that for the core of the beam (smaller amplitude values).

\section{Summary}
The KDE technique has been used to characterize the MICE muon beam. Its application yielded a detailed understanding of the beam phase-space density and its evolution through a solenoidal cooling lattice. The density of the muons in the core of the beam has been shown to increase, which is an unambiguous sign of cooling.

In general, the KDE approach helps with finding a surface characterizing any given fraction of the beam very precisely and can lead to its application in nonlinearity studies of other accelerators (e.g. IOTA at Fermilab \cite{ref21}).  

The change in density using the KDE technique has proven to be robust against beam loss and nonlinear effects in the MICE beam. Application of KDE to MICE data is in progress. 

\section{Acknowledgements}
MICE is supported by U.S. Department of Energy (DOE), U.K. Science and Technology Facilities Council (STFC), Instituto Nazionale del Fisica Nucleare (INFN), the Japan Society for the Promotion of Science, and the Swiss National Science Foundation, in the framework of the SCOPES program. 

We are thankful to Daniel Kaplan, Victoria Blackmore, and J. Scott Berg for valuable discussions. The work presented here has been supported by the U.S. Department of Energy (DOE) Office of Science Graduate Student (SCGSR) program under contract No. DE-AC05-06OR23100 during the 2015 Solicitation 2 award period.

\end{document}